\documentclass[apj]{emulateapj}
\usepackage{amssymb,amsmath}

\shorttitle{Mapping reionization with Ly$\alpha$ emission}
\shortauthors{Cantalupo, Porciani \& Lilly}
\slugcomment{To be published in the Astrophysical Journal}

\newcommand{\ergscmarcsec}{$\mathrm{erg}\mathrm{s}^{-1}\mathrm{cm}^{-2}\mathrm{arcsec}^{-2}$}

\newcommand{\lya}{Ly$\alpha$\ }
\newcommand{\lyam}{\mathrm{Ly}\alpha }

\newcommand{\mr}{\mathrm}

\begin{document}

\def\simlt{\mathrel{\rlap{\lower 3pt\hbox{$\sim$}} \raise
        2.0pt\hbox{$<$}}} \def\simgt{\mathrel{\rlap{\lower
        3pt\hbox{$\sim$}} \raise 2.0pt\hbox{$>$}}}
\def\simgt{\mathrel{\rlap{\lower 3pt\hbox{$\sim$}} \raise
        2.0pt\hbox{$>$}}} \def\simgt{\mathrel{\rlap{\lower
        3pt\hbox{$\sim$}} \raise 2.0pt\hbox{$>$}}}


\title{Mapping Neutral Hydrogen during Reionization \\
 with the \lya Emission from Quasar Ionization Fronts} 

\author{Sebastiano Cantalupo, Cristiano Porciani and Simon J. Lilly}
\affil{Institute for Astronomy, ETH Z\"urich, CH-8093 Z\"urich, Switzerland}
   
\email{cantalupo@phys.ethz.ch}


\begin{abstract}

We present a new method
to directly map the neutral-hydrogen distribution 
during the reionization epoch
and to constrain the emission properties 
of the highest-redshift quasars (QSOs).
As a tracer of HI, we propose to use
the \lya radiation produced by quasar ionization fronts (I-fronts)
that expand in the partially ionized intergalactic medium (IGM)
before reionization is complete.
These \lya photons are mainly generated by collisional excitations
of hydrogen atoms in the boundary of the rapidly expanding HII region.
The observable signal is produced 
by the part of the I-front 
that lies behind the QSO with respect to the observer.
The expected \lya flux depends 
on the properties of both the QSO (e.g. its total ionizing rate and
spectral hardness) and the surrounding medium (e.g. the local density
and the mean hydrogen neutral fraction).
Combining two radiative transfer models 
(one for the QSO ionizing radiation and one for the 
\lya photons), we estimate the expected
\lya spectral shape and surface brightness (${\rm SB}_{\lyam}$) 
for a large number of configurations where we varied both the properties
of the ionizing QSO and of the surrounding medium.
We find that the expected signal is observable as a 
single (broad) emission line with a characteristic width of 
$100-200$ km s$^{-1}$. 
The expected ${\rm SB}_{\lyam}$ produced at redshift $z\simeq 6.5$ 
within a fully neutral region (at mean density)
by a typical QSO I-front
lies in the range $10^{-21}-10^{-20}$ erg s$^{-1}$ cm$^{-2}$ arcsec$^{-2}$
and decreases proportionally to $(1+z)^2$ for a given QSO age.
QSOs with harder spectra may produce a significantly brighter emission
at early phases.
The signal may cover up to a few hundred square arcmin
on the sky
and should be already detectable with current
facilities by means of moderate/high resolution spectroscopy.
The detection of this \lya emission can shed new light on the 
reionization history, the age and the
emission properties of the highest-redshift QSOs.

\end{abstract}


\section{Introduction}

The epoch of reionization (EoR) marks the time at which 
the first luminous sources ionized the 
mostly neutral
intergalactic medium (IGM).
Uncovering when and how the EoR took place 
represents one of the most fundamental questions of modern cosmology
and considerable efforts 
are being made to understand this era.
 
The recent polarization measurements of the cosmic microwave background (CMB)
by the WMAP satellite suggest that the universe was mostly
neutral at redshift $z\simgt14$ (Page et al. 2006).
On the other hand, at $z<5$,
hydrogen absorption features in the spectra of luminous quasars
are resolved into individual \lya forest
lines (e.g. Rauch 1998) as expected in a highly ionized IGM.
At $z\gtrsim6$, quasar spectra start showing complete Gunn-Peterson 
absorption throughs (GPT) which might 
suggest a rapid increase in the neutral fraction of cosmic hydrogen
(Becker et al. 2001; Fan et al. 2003; White et al. 2003).  
However, the \lya transition has a large cross-section and 
the presence of GPTs only requires very low values for 
mean neutral fraction of cosmic hydrogen: 
$\langle x_{HI}\rangle \gtrsim10^{-3}$
(Fan et al. 2002; Songaila 2004; Oh \& Furlanetto 2005). 

Stronger constraints on $\langle x_{HI}\rangle$ 
can be obtained from the detailed spectral shape of the GPT, i.e.
from the size of the HII region produced by the QSO itself (White et al. 2003; 
Mesinger \& Haiman 2004; Wyithe \& Loeb 2004; Yu \& Lu 2005; Fan et al. 2006; 
Maselli et al. 2006; Bolton \& Haehnelt 2007).
The physical scales of these regions are typically inferred to be $\sim5$ Mpc
at $z>6$ (Fan et al. 2006), an order of magnitude larger than the HII bubbles 
expected from clustered star-forming galaxies 
(Furlanetto, Zaldarriga \& Hernquist 2004). 
The size of the HII region around a QSO depends
on the neutral density of the surrounding medium but also
on the source luminosity and age (more specifically, on its light curve).
With some knowledge of the QSO parameters 
it is then possible to 
constrain the value of $\langle x_{HI}\rangle$. 
Whythe et al. (2005) applied this method to
seven QSOs at $6\lesssim z\lesssim 6.42$ and derived
$\langle x_{HI}\rangle\gtrsim0.1$. 
Similarly, from the size of the \lya and Ly$\beta$ troughs of a QSO at 
$z=6.28$, Mesinger \& Haiman (2004) found $\langle x_{HI}\rangle\gtrsim0.2$.
On the contrary, Fan et al. (2006) did not find strong evidence
for a mean hydrogen neutral fraction as high as 0.1 at $z=6.4$.
Apart from systematics, the main uncertainty of this method
lies in the estimate of the QSO age.
As a matter of fact, the current observed sizes seem to be equally consistent
with both a significantly neutral and a highly ionized surrounding IGM 
(e.g. Bolton \& Haehnelt 2007). 

Additional (indirect) information
about the ionized fraction of the IGM 
can be obtained from the luminosity-function evolution
of \lya emitting galaxies (LAE) at $z>6$ 
(Miralda-Escud\'e 1998; Madau \& Rees 2000; Haiman 2002; 
Santos 2004; Taniguchi et al. 2005; Furlanetto et al. 2006; Malhotra \& Rhoads 2006; Kashikawa et al. 2006), 
and from their clustering properties (McQuinn et al. 2007).
The key idea is that the observed \lya flux from high-$z$ galaxies 
should strongly depend on the neutral fraction of their surrounding
medium. 
However, these observations are challenging and difficult to interpret.
Any change in the luminosity function due to the EoR must 
be distinguished from an intrinsic physical evolution of the sources, 
as well as the effects of dust or galactic winds. 
Similarly, clustering signatures of the EoR can only be detected
by comparing the spatial distribution of a given galaxy population
(e.g. Lyman-break galaxies) with that of the sub-sample of LAEs
(see Fan, Carilli \& Keating 2006 for a recent review).

A very promising method for studying the EoR
is the detection of the redshifted 21-cm emission from neutral hydrogen 
(Madau, Meiksin \& Rees 1997; see Furlanetto, Oh \& Briggs 2006 
for a recent and exhausitive review).
This technique has received a lot of attention recently as a concrete
possibility of studying the detailed history of the EoR.
From the technical point of view, however, it requires
an extremely challenging measure:  
the intrinsic 21-cm signal, detected as a 
brightness-temperature fluctuation against the CMB, is at least four orders of 
magnitude smaller than the emission from our Galaxy. 
Other important foregrounds
include radio recombination lines, terrestrial interference and ionospheric 
distortion. 
The hope is to separate the fluctuating spectral features due to
21-cm emission from the smooth spectrum of the foregrounds.
Anyway, 
given the difficulties of high signal-to-noise imaging, attention has been 
focussed on statistical measurements, like the power-spectrum analysis. 
First-generation facilities (as LOFAR and MWA, now under construction)
should only detect a statistical signature of reionization
For a direct detection of the three-dimensional structure of HI and HII
regions (the so called  ``21-cm tomography'') 
we will probably have to wait for a subsequent generation of radio arrays with 
much larger collecting area (e.g. SKA).

Are there other possibilities to directly detect the high-z IGM in emission?
\lya is the strongest hydrogen emission line. 
This transition (from the 2P to the 1S level of
atomic hydrogen) has a spontaneous emission coefficient 
$A_{21}=6.25\times10^{8}$ s$^{-1}$. 
An efficient mechanism to populate the $n=2$ level, with consequent
\lya emission, is the HII recombination process. 
Unfortunately, the recombination rate is
typically very low and recombinaton radiation from optically thin hydrogen in the 
high-redshift IGM cannot be detected with current instruments
(see e.g. Hogan \& Weymann 1987; 
but also Baltz, Gnedin \& Silk 1998).

Optically thick and self-shielded HI clouds exposed to strong UV radiation are 
expected to re-emit a significant
part of the impinging flux in the form of ``fluorescent'' \lya emission 
(Gould \& Weinberg 1996; 
Cantalupo et al. 2005). A recent attempt to detect fluorescent emission, 
and thus HI protogalactic-clouds,
at intermediate redshift ($z\sim3$) has provided a number of promising
candidates (Cantalupo, Lilly \& Porciani 2007).
However, self-shielded clouds correspond to overdense 
regions with relatively small sizes that trace just a small
fraction of the neutral hydrogen distribution.

Under favourable conditions,
the excitation of atomic hydrogen due to collisions with energetic electrons 
populates the $n=2$ level in a much more efficient way than recombinations.
In this paper, we show that the corresponding \lya emission can be 
efficiently used to map the neutral hydrogen distribution during the EoR.
This emission should be already detectable with current observational
facilities and could be used to shed light on  
the bulk of neutral hydrogen during the EoR. 

The layout of the paper is as follows. 
In \S 2 we summarize the basic physics of \lya emission from 
recombination and collisional excitation processes. 
In \S 3 we show that collisional excitations should efficiently produce 
\lya emission within the ionization fronts of high-redshift QSOs.
We also explain how 
it is possible to map the
distribution of neutral hydrogen behind the QSO using this emission.
In \S 4 we present 
more sophisticated models of \lya emission based on numerical simulations
including a detailed treatment of radiative transfer.
In \S 5 we discuss the dependence of our results on model parameters and 
possible detection strategies. 
Finally, in \S 6, we summarize our results.
The reader who is not interested into the technical details of the model
can safely read \S 5.2 right after \S 3 (focussing also on Figures 7 and 8). 

Throughout this paper, we adopt a standard, flat $\Lambda$CDM cosmological 
model with mass-density parameter $\Omega_{\rm m}=0.24$, a baryonic contribution
of $\Omega_{\rm b}=0.044$, and a present-day Hubble constant 
of $H_0=100\,h$ km s$^{-1}$ Mpc$^{-1}$  with $h=0.73$. 
We also adopt the notation ``pMpc''  
to indicate proper distances measured in Mpc. 


\section{The physics of Ly$\alpha$ emission}

 There are two main channels that drive the production of Ly$\alpha$ photons in the IGM:
i) recombination processes and ii) collisional excitations by free electrons. 
Both mechanisms directly arise from the ionization of the hydrogen atom.

\subsection{\lya from recombinations}

When an electron is captured by
a free proton, it can either directly populate the ground 1 $^2S$ level 
(with the emission of a Lyman continuum photon) 
or an excited state from which it cascades by downward transitions to the 
1 $^2S$ level 
(with the emission of a continuum photon plus several transition lines). 
Cascades that populate the 2 $^2P$ state decay to the 1 $^2S$ level 
by emitting a Ly$\alpha$ photon, while
cascades to the 2 $^2S$ level subsequently decay via two-photon emission.

\begin{figure}
\plotone{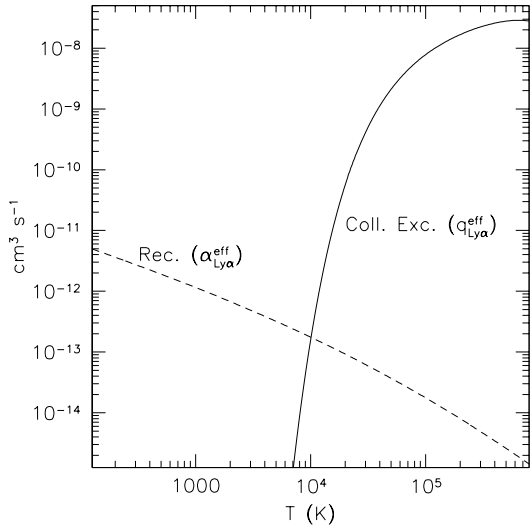}
\caption{Effective Ly$\alpha$ production rate from recombination processes ($\alpha^{eff}_{Ly\alpha}$, 
dashed line) and collisional excitations ($q^{eff}_{Ly\alpha}$, solid line) 
as a function of the gas (electronic) temperature.}
\label{f1}
\end{figure}

Electrons that populate $n\geq3$ levels may also decay to the ground state 
with the emission of
another Lyman-series line (with no two-photon or Ly$\alpha$ emission).  
However, even in the low-density IGM, 
the typical Lyman-line emitting regions have quite large optical
depths ($\tau$) in the Lyman resonance lines (Osterbrock 1989).  
Therefore, a good approximation 
(unless the medium is fully ionized) is to assume that every Lyman-line
photon from $n\geq3$ levels is converted 
into lower-series photons plus either Ly$\alpha$ or two-photon
radiation (Case B approximation, Baker \& Menzel 1938). 
In this case, it is convenient to write the effective 
Ly$\alpha$ emission coefficient from recombinations as:
\begin{equation}
\alpha^{\mr{eff}}_{\lyam}(T)=\epsilon^{\mr{B}}_{\lyam}(T) \alpha_{\mr{B}}(T)\ ,
\end{equation}
where $\alpha_{\mr{B}}(T)$ is the hydrogen total recombination coefficient excluding recombinations to the
ground level, 
and $\epsilon^{\mr{B}}_{\lyam}(T)$ is the fraction of those recombinations 
producing \lya photons.
Combining the tabulated values by Pengelly (1964; for $T>10^3\,\mathrm{K}$) 
and Martin (1988; for $T<10^3\,\mathrm{K}$), 
we have derived the following fitting formula for $\epsilon^{\mr{B}}_{\lyam}(T)$ 
(accurate to the 0.1\% in the temperature range 
$100\,\mathrm{K}<T<10^5\,\mathrm{K}$):
\begin{equation}
\epsilon^{\mr{B}}_{\lyam}(T)=0.686-0.106\log(T_4)-0.009\cdot(T_4)^{-0.44}\ ,
\end{equation}
where $T_4=T/10^4\,\mathrm{K}$. 
Note that the value of $\epsilon^{\mr{B}}_{\lyam}(T)$ varies very little
with temperature, ranging beetwen 0.68 and 0.61 for $T_4=1-5$. 
The temperature dependence of $\alpha^{\rm eff}_{\lyam}(T)$
is shown in Figure \ref{f1} as a dashed line.  
Note that the value of $\epsilon^{\mr{B}}_{\lyam}(T)$ varies very little
with temperature, ranging beetwen 0.68 and 0.61 for $T_4=1-5$. 

 The volume \lya emissivity due to radiative recombinations is given by:
\begin{equation}
\frac{4\pi j_{\lyam}}{h\nu_{\lyam}}=n_{\mr{e}} n_{\mr{p}} \alpha^{\mr{eff}}_{\lyam}
\end{equation}
where $j_{\lyam}$ is the emissivity (energy radiated per unit time, volume and
solid angle),
$h\nu_{\lyam}=10.2$ eV is the energy of a \lya photon and $n_e$, $n_p$ are, respectively, the
electron and proton number densities. Note that the coefficient
 $\alpha^{\mr{eff}}_{\lyam}$
is independent from the assumption regarding 
the fate of Lyman-continuum photons. 
Wherever these photons are absorbed, and also in the case they escape 
the medium, 
the local values of $n_e$ and $n_p$ will change accordingly, modifying the \lya emissivity.

\subsection{\lya from collisional excitations}

Free electrons can interact one or
more times with neutral hydrogen atoms before eventually recombining. 
Given the relatively low densities of the IGM, 
we can imagine the free electrons colliding with
hydrogen in the ground state and transferring to it a fraction of their energy. %
If this energy (i.e. the electron temperature) is high enough, the collision
will excite the atomic levels with subsequent decay via line emission. 
The efficiency of the process is strongly temperature dependent. In particular, the collisional
excitation coefficient for the transition from the ground level (1) to the level $nl$ is given by:
\begin{equation}
q_{1,nl}=\frac{8.629\times10^{-6}}{\sqrt{T}}\frac{\Omega(1,nl)}{\omega_1} e^{-E_{1,n}/k_BT} \mathrm{cm^3 s^{-1}}\ ,
\end{equation}
where $\Omega(1,nl)$ is the temperature dependent effective collision strength, $\omega_1$ is the statistical
weight of the ground state and $E_{1,n}$ is the energy difference beetwen the ground and the $nl$ level.
We define the effective collisional excitation coefficient for 
\lya emission as:
\begin{equation}
q^{\mr{eff}}_{\lyam}=q_{1,2p}+q_{1,3s}+q_{1,3d}
\end{equation}
where we consider excitation processes up to the level $n=3$ that will 
eventually produce \lya radiation (in Case B 
approximation). 
The contribution from higher levels is completely negligible for the
electron temperatures we are interested in.
Similarly, we neglect angular-momentum-exchanging processes 
from proton and electron 
collisions (e.g. the 2s-2p and 2p-2s transition) that, in principle, might 
boost or suppress \lya emission. The typical densities of the IGM are
too low for these transitions to be important (Osterbrock 1989).
 We use the polynomial fits of Giovanardi, Natta \& Palla (1987)
 to compute $q^{\mr{eff}}_{\lyam}$.
 The result is shown as a solid line in Figure \ref{f1}.

 The total \lya emissivity coming from
radiative recombinations and collisional excitations is thus given by:
\begin{equation}\label{lya_em}
\frac{4\pi j_{\lyam}}{h\nu_{\lyam}}=n_{\mr{e}} n_{\mr{p}}\ \alpha^{\mr{eff}}_{\lyam} + 
 n_{\mr{e}} n_{\mr{HI}}\ q^{\mr{eff}}_{\lyam}\ ,
\end{equation}
where $n_{\mr{HI}}$ is the neutral hydrogen density.
Higher ionization fractions favour \lya emission from 
radiative recombinations, while
higher temperatures increase the contribution from collisional excitations 
provided that a significant fraction of hydrogen atoms remain neutral.
Therefore,
collisional excitation significantly contributes to the total \lya 
emissivity in relatively hot regions where $n_{\mr{e}}\sim N_{\mr{HI}}$,
i.e. where the neutral fraction ($x_{\mr{HI}}$) is about 0.5. 
In particular, if the temperature of these regions
is in the range $2-5\times10^4\,\mathrm{K}$, collisional excitation may 
increase the total \lya signal of several orders
of magnitudes with respect to radiative recombinations.

\section{\lya emission from quasar ionization fronts during reionization}\label{sec3}

 The two requirements for efficient \lya production via collisional excitation
($T\sim$ a few $\times 10^4\,\mathrm{K}$ and $x_{\mr{HI}}\sim0.5$) 
are met within the ionization fronts (I-fronts) 
produced by powerful UV sources with a hard spectrum (like QSOs)
located in a mostly neutral IGM. 
An I-front constitutes the transition region between the inner HII zone 
and the outer (predominantly neutral) IGM.
In this transition layer, 
the hydrogen neutral fraction varies by several orders of magnitudes
within a length scale determined by the mean free path of the ionizing photons. 
The thickness of the I-front is generally negligible with respect to the characteristic 
radius of the HII region.
Sources with harder spectra produce thicker I-fronts 
(for a given IGM density)
because of the frequency dependence of the photo-ionization cross-section.
If the I-front is not in photo-ionization equilibrium, 
ionization fractions evolve with time
and the transition zone moves outward (the HII region expands).
The propagation of the I-front produced by powerful UV sources like QSO 
can be highly relativistic, racing ahead of the hydrodynamic response of the 
ionized gas for the entire lifetime of the source (Shapiro \& Giroux 1987).

\subsection{A simple estimate of the \lya signal}

As shown in equation (\ref{lya_em}),
the volume emissivity of \lya radiation within the I-front depends 
on the electron, proton and HI densities, and on the local temperature.
Detailed modelling of these quantities within an expanding I-front
requires an accurate 
numerical solution of the radiative transfer problem and is 
postponed to the next Section.
However, we can give an order-of-magnitude estimate of the expected 
amplitude of the signal following a simple analytical reasoning. 

Let us assume that a bright QSO turns on at redshift $z_{\mr in}>6$, 
before the reionization process is complete, 
and produces a rapidly expanding I-front that completely ionizes 
the surrounding IGM. 
Let us further assume, for simplicity, that the IGM is at mean cosmic
density, the hydrogen neutral fraction 
varies linearly (from 0 to 1) with distance within the 
thickness of the I-front, and that the I-front temperature is uniform. 
 The actual value of the I-front temperature depends on the detailed balance
between photoheating and cooling processes 
(see, e.g. Miralda-Escude \& Rees 1994). In the following we
estimate the expected Ly$\alpha$ signal for a given I-front temperature
and medium density
\begin{footnote}{
The appropriate scaling relations based on other parameters, like
the QSO age, ionizing rate and local overdensity will be derived in \S 5.}
\end{footnote}.
Typical values for this temperature can be inferred from 
the far-UV spectral indices of QSOs (see e.g., Telfer et al. 2002)
and lie in the range
$2-4\times10^4$ K if the IGM is at mean cosmic density 
(see e.g., Abel \& Haehnelt 1999).
At these temperatures
we can neglect \lya emission due to radiative recombinations and 
write the integrated \lya flux at the inner edge of the I-front as:
%
\begin{eqnarray}\label{I1}
\lefteqn{I_{\lyam} \simeq\frac{1}{\pi}\int_{\mr{I-front}} \!\!\!\!\!n_{\mr{HI}}(r)n_{\mr{e}}(r)q^{\mr{eff}}_{\lyam}(T)\mathrm{d}r}\nonumber \\
          & & \simeq\frac{1}{\pi} n^{2}_{\mr{H}}\, C \,\chi_{\mr{e}}\, 
\left(\frac{S}{6}\right)\, q^{\mr{eff}}_{\lyam}(T)\ ,
\end{eqnarray}
%
where  $C\equiv \langle n_{\mr{H}}^2\rangle/\langle n_{\mr{H}}\rangle$ is the hydrogen 
clumping factor, $\chi_{\mr{e}}$ the 
factor that accounts for the contribution of ionized helium to the electron 
density, and $S$ denotes
the thickness of the I-front. 
The factor $1/6$ derives from the integration of $x_{\mr{HI}}(1-x_{\mr{HI}})$
over the front, while the factor $1/\pi$ accounts for the angular distribution
of the emitted photons (Gould \& Weinberg 1996).
The actual value of $S$ depends on several
factors, including the spectrum of ionizing radiation $F_\nu$ 
and the local density.
As a first-order approximation, we can assume that the
size of the I-front is given by the mean free path 
of the ionizing photons with 
mean frequency $\langle \nu \rangle=\int_{\nu_0}^\infty F_\nu\, d\nu
/\int_{\nu_0}^\infty (F_\nu/h_{\mr P}\nu)\, d\nu$
(with $\nu_0$
the hydrogen ionization threshold and $h_{\mr P}$ the Planck constant), 
i.e. $S\simeq(n_{\mr{H}} \sigma_{\langle \nu \rangle })^{-1}$, where
$\sigma_\nu$ is the (frequency dependent) photo-ionization cross-section.
Substituting $S$ in equation (\ref{I1}), 
replacing the hydrogen density with $n_{\mr{H}}=n_{\mr{H},0}(1+z)^3$
 (where $n_{H,0}$ is the comoving mean number density), and including the redshift dimming,
we obtain the observed integrated surface brightness (SB) in photons:
\begin{equation}
\Phi_{\lyam}\simeq\left(\frac{f_{\mr{esc}}\,C\,\chi_{\mr{e}}}{6\pi}\right)n_{\mr{H},0}\frac{q^{\mr{eff}}_{\lyam}(T)}
{\sigma_{\langle \nu \rangle }}
\end{equation} 
where $f_{\mr{esc}}$ is the fraction of \lya photons that manage to escape 
from the I-front along the line of sight.
Remarkably, the observed SB (in photon number) does not depend 
on the QSO redshift for a given temperature of I-front. 

How strong is this signal?
The exact value of $f_{\mr{esc}}$ is difficult to estimate as it depends on 
several factors, including
the size of the HII region and the value of the residual HI in proximity of 
the I-front. 
Detailed calculations (see \S\ref{secMod}) suggest that
$f_{\mr{esc}}\sim0.5$.
Typical values of the clumping factor at $z>6$, estimated
from simulations (e.g., Gnedin \& Ostriker 1997), are of order of 
$C\sim30$.
Therefore, assuming $\langle \nu \rangle\simeq3\,\nu_0$, $\chi_{\mr e}=1.2$ 
and fixing the temperature to $T\simeq3\times10^4$ K, we obtain:
\begin{equation}
\Phi_{\lyam}\simeq 200\ \mathrm{photons}\ 
\mathrm{s}^{-1}\ \mathrm{cm}^{-2}\ \mathrm{sr^{-1}}\ ,
\end{equation}
at the top of Earth's atmosphere, or, equivalently:
\begin{equation}
{\mr SB}_{\lyam}\simeq 10^{-20}\ 
\mathrm{erg}\ \mathrm{s}^{-1}\ \mathrm{cm}^{-2}\ \mathrm{arcsec^{-2}}\ ,
\end{equation}
for a QSO at $z\sim 6.5$.

 Although much fainter than the limit fluxes of present-day surveys for \lya emitters
at $z\sim6.5$ ($\sim 10^{-18}\mathrm{erg}\ \mathrm{s}^{-1}\ \mathrm{cm}^{-2}\ $
for apertures of order of few arcsec$^2$; Tran et al. 2004; Kashikawa 
et al. 2006), the emitting region can cover several hundred 
comoving Mpc$^2$. This corresponds to several hundred arcmin$^2$
on the sky at $z\sim6.5$ and makes the signal detectable with current
facilities by means of moderately-high resolution spectroscopy.
\footnote{
The detectability of this signal with present and future instruments will be 
discussed in \S 5.2.} 

Observations of this emission would provide direct evidence for 
the presence of an I-front, 
shedding light on the properties of both the ionizing source and the
surrounding medium. 
In particular, as we will show in \S 5,
the angular shape of the apparent I-front can constrain the
QSO age and its angular emission properties.

\begin{figure}
\epsscale{1.1}
\plotone{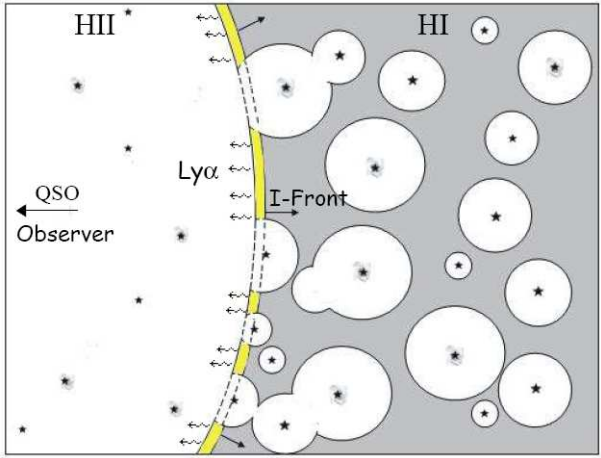}
\caption{This schematic cartoon illustrates how to map the HI distribution using the \lya emission
from the I-front of an high-z QSO.
The QSO and the observer are on the left side. The ionizing photons, in the plane
of the figure, propagate from left to right and create an I-front 
wherever they encounter
a partially neutral region. Within the I-front, the interactions between  
neutral hydrogen and the energetic electron released by photo-ionization
produce \lya photons via collisional excitation (yellow stripes). The \lya emission escapes
the I-front trough the HII region in the direction of the observer. 
The regions already ionized by local sources
(bounded by the dashed lines) do not produce \lya via collisional 
excitations and
will appear as holes in the 2-D \lya map, similarly to a 21-cm tomography.
This strictly applies to a young QSO with a relativistically expanding
I-front. For older quasars, the entire boundary of the HI region can emit
\lya photons.}
\label{f2}
\end{figure}

\subsection{\lya map of the HI distribution during reionization}

Detecting the \lya emission from the I-front makes it possible to
directly map the distribution of neutral hydrogen 
behind the QSO at the location of the I-front.
In fact,
the distribution of the \lya signal on the sky
traces the local density of HI as it was before the arrival of the I-front. 
We will illustrate this with a simple example.
Let us consider two
patches of the IGM (at mean density) that are simultanously reached
by the QSO I-front. Right before the arrival of the front, one of them
is still neutral ($x^{\mr a}_{\mr{HI}}=1$) and the other is
significantly ionized ($x^{\mr b}_{\mr{HI}}<0.1$) 
by a local faint source.
Let us further assume that the local source in the second region 
has a softer spectrum than a QSO and thus that
the (initial) temperature of the free electrons is low, $T\lesssim10^4$ K.
While the I-front crosses the two regions,
energetic electrons are released 
by hydrogen atoms
and collisional excite the remaining neutral hydrogen. 
The \lya emissivity reaches a maximum at the center of the I-front where the 
actual neutral fraction is about $0.5 x_{\mr{HI}}$
and the temperature peaks. 
However, the mean energy  
of the free electrons and the number density of neutral-hydrogen atoms 
in the second region 
are both reduced by a factor 
$x^{\mr a}_{\mr{HI}}/x^{\mr b}_{\mr{HI}}>10$ with respect to the first region:
the \lya emissivity is thus reduced by more than a factor of 100.

A schematic description of how this effect can be used to map the HI distribution  
is shown in Figure \ref{f2} for a uniform IGM. The locally ionized
regions will appear as ``holes'' in the 2-dimensional \lya map.
\footnote{This strictly applies to young QSOs with I-fronts that are expanding 
at ultrarelativistic speeds. For older QSOs, the front will move faster in
the regions pre-ionized by other local sources and will produce bright \lya
emission from the entire (irregular) boundary of the HI region.}
Given the large angular extension of the HII region, mapping the \lya emission from 
the I-front behind a high-redshift QSO can 
effectively constrain the topology and the history of the reionization
process.
 
\section{Modelling the \lya signal}\label{secMod}

The rate equations that regulate the temperature and  
the ionization-fraction profiles inside the expanding I-front cannot be solved 
analytically. 
These quantities depend
on the local ionizing spectrum that is determined by non-local
radiative-transfer (RT) effects. 
For instance, higher energy photons will be absorbed at higher
column densities, causing an effective hardening of the ionizing spectrum 
at larger distances from the source.
Therefore, detailed modelling of the \lya emissivity 
requires a full RT transfer calculation for the ionizing radiation.

\subsection{Continuum Radiative Transfer}

To follow the radiative transfer of the continuum ionizing radiation
from the QSO,
we use a three-dimensional, photon-conserving, and time-dependent 
code based on a ray-tracing algorithm
(Cantalupo \& Porciani, in preparation). 
This code has been developed to study
the propagation of I-fronts in cosmological simulations and it
includes an adaptive refinement scheme of the computational grids 
(in space and time) on the front.
Among other features, it accounts for the presence of
multiple ionizing sources and for the propagation of ionizing radiation
produced by recombinations. 
The time-dependent rate equations include the 
evolution of HI, HeI and HeII. 
The gas temperature is computed including the
energy input due to
photoionizations and collisional ionizations of 
the three species and all the relevant
cooling processes (recombinations, collisional excitations/ionizations, dielectronic recombination of HeII,
bremsstrahlung, Compton and Hubble cooling). 

For simplicity, in the present study, we consider a single (steady) ionizing QSO 
surrounded by a uniform (cosmologically expanding) medium.
We follow the evolution of the I-front over a few $10^8$ yr,
i.e up to distances of
70-140 comoving Mpc from the source (depending on
the QSO luminosity). 
The adaptive mesh refinement (AMR) scheme 
allow us to resolve the I-fronts with a large number of 
(optically-thin) cells of a few comoving kpc at all times.
A more general configuration based on the density and velocity fields
extracted from a hydro-dynamical simulation will be analyzed in future work.

\subsection{Finite light-travel time effects}

High-redshift QSOs are expected to produce 
relativistically expanding I-fronts for an important fraction of their lifetime 
(Shapiro et al. 2006).
This is difficult to follow with most RT codes
(see Iliev et al. 2006 for a recent compilation),
including our own,
in which the ray-tracing algorithm assumes an infinite light speed and
ionization fronts tend to propagate faster than light at early times.
A quick fix is obtained by limiting the ray-tracing up to a maximum distance
but this would lead to the loss of photon conservation.
We thus decided to follow a novel approach.
For a steady ionizing source in a uniform medium,
the correct speed of the I-front at a given radius, $v_{\mr I}(r_{\mr I})$, 
can be expressed in terms of its counterpart obtained assuming
an infinite speed of light, $v_{\mr I,NR}(r_{\mr I})$, by the relation
\begin{equation}
v_{\mr I}(r_{\mr I})=\frac{v_{\mr I,NR}(r_{\mr I})}{1+v_{\mr I,NR}(r_{\mr I})/c}\ ,
\end{equation}
regardless of the difference in physical times at which the I-front reaches 
the radius $r_{\mr{I}}$ (White et al. 2003; Yu 2005; Shapiro et al. 2006).
Therefore, we can get the correct velocity of the I-front 
by redefining the time variable as follows.
We associate to each time step used in the radiative transfer code
$\Delta t_{\mr{NR}}$ an actual time step $\Delta t_{\mr{R}}$
simply given by:
\begin{equation}
\Delta t_{\mr{R}}=\Delta t_{\mr{NR}} (1+v_{\mr{I},\mr{NR}}/c)\ .
\end{equation}
Note that this is an approximate solution, since the optical
depth of the photons is still computed at the time of their emission
and does not take 
into account the changes taking place in the medium during their
propagation time.

\begin{figure}[t!]
\epsscale{0.9}
\plotone{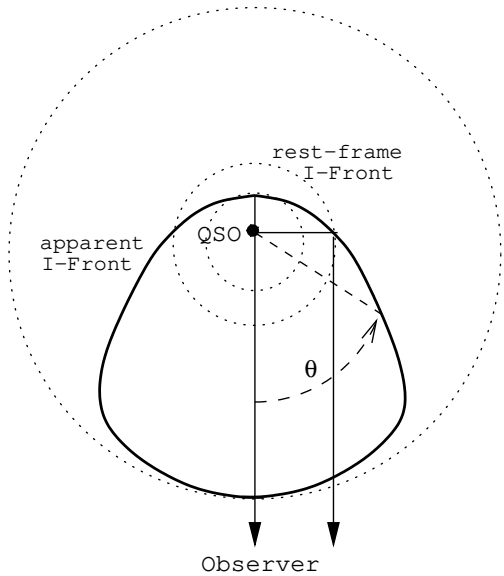}
\caption{Schematic view of the apparent I-front shape (solid line) around an isotropically emitting QSO.
The dotted circles represent the I-fronts 
in the QSO rest-frame at different epochs. The departure from the
spherical shape is due to the relativistic expansion speed of the I-front c
ombined with finite light-travel effects.
The apparent shape is rotationally symmetric along the quasar line of sight.  
The positions of the apparent and rest-frame I-front coincide for 
$\theta=\pi/2$, while the apparent
expansion of the I-front at $\theta=0$ is superluminal.}
\label{fig3}
\end{figure}

\begin{figure}[t!]
\epsscale{1.2}
\plotone{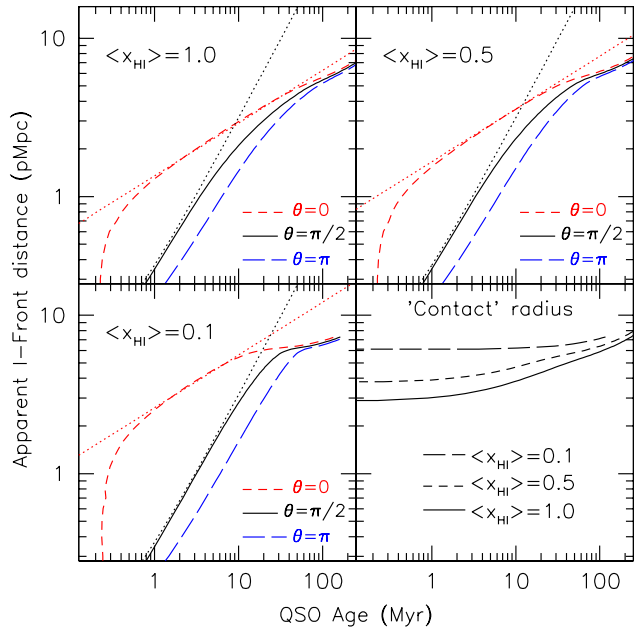}
\caption{Time-evolution of the apparent I-front 
distance from the QSO (in physical Mpc)
as a function of the angle $\theta$ (defined as in 
Figure \ref{fig3}) and the mean neutral fraction of the environment 
$\langle x_{\mr{HI}}\rangle$ (measured before the arrival of the I-front)
obtained with our radiative transfer code 
including finite light-travel effects.
Dotted lines are for reference and indicate propagation at 
the speed of light (black) and $\propto t^{1/3}$ (red).
The bottom right panel shows the evolution of the ``Contact'' radius, 
i.e. the position where the I-front 
closest to the observer ($\theta=0$) is reached by 
a photon emitted by the I-front behind the quasar ($\theta=\pi$). 
Note that the apparent $\theta=\pi/2$ and the
rest-frame I-front expansions are identical.
 All the panels assume a QSO that turns on at $z_{in}=6.5$
with total ionizing rate $\dot{N_{\gamma}}=10^{57}$ ph s$^{-1}$, far-UV spectral index $\alpha=-1.7$, and a clumping factor $C=35$. 
These parameters are compatible with observed QSOs showing
a Gunn-Peterson trough at $z>6$ (Yu \& Lu 2005).}
\label{f4}
\end{figure}

\subsection{The apparent evolution of the I-front}

The observed shape of the I-front 
(i.e. the relation between the QSO age, $t$, and the I-front distance 
from the QSO, $r_{\mr I}(t)$, as a function of the viewing angle $\theta$ defined
in Figure 3)
will differ from the proper shape because of finite 
light-travel effects (see equations (1)-(3) in Yu 2005).
In fact,
the photons we receive from the part of the I-front that is
moving towards us travel a smaller distance 
than the photons emitted behind the quasar. 

In Figure \ref{f4}, we show
the time-evolution of the apparent I-front position
for three different values of 
$\theta$ and $\langle x_{\mr{HI}}\rangle$ (the mean neutral fraction
of the medium surrounding the ionization source before the latter turns on).
This is the output of a simulation where a
luminous QSO (with a total ionizing rate $\dot{N_{\gamma}}=10^{57}$ ph s$^{-1}$,
and far-UV energy spectral index $\alpha=-1.7$)
turns on at $z_{\mr in}=6.5$ in an uniform medium at mean density 
with a clumping factor $C=35$.
This parameter set is consistent with observations of QSOs 
showing a GPT at $z>6$ (Yu \& Lu 2005).
From now on, the evolution of the I-front 
in a fully neutral 
medium ($\langle x_{\mr{HI}}\rangle=1$) 
surrounding this source will constitute our reference model.

The time-evolution of the I-front position obtained from our code 
is in good agreement with the approximate analytical solution by Yu (2005) 
and Shapiro et al. (2006).
For a given QSO ionizing rate, the apparent position of the I-front closest
to the observer (red short-dashed line in Figure 4) is very sensitive
to both the QSO age and $\langle x_{\mr{HI}}\rangle$. This is not the case for
the apparent position of the \lya emitting I-front \emph{behind} the QSO
(blue long-dashed line in Figure 4),
that determines fairly well the QSO age (at least for $\langle x_{\mr{HI}}\rangle\gtrsim 0.1$).
Combining the information from the \lya emitting I-front and the
position of the GPT in the QSO spectrum, we can significantly constrain both the
QSO age and $\langle x_{\mr{HI}}\rangle$.

At any given time, it is useful to define the
``contact'' radius ($R_{\mr{t}}$) as
 the position where the I-front closest to the observer ($\theta=0$) 
is reached by 
the photons emitted by the I-front behind the quasar ($\theta=\pi$).
This quantity determines the strength of the damping-wing absorption
for the observed \lya signal from the I-front at $\theta=\pi$.
The evolution of the contact radius for our reference model is shown in 
the bottom right panel of Figure \ref{f4}.
Note that $R_{\mr{t}}$ keeps nearly constant at early times and grows afterwards.
Therefore, even when the rest-frame radius of the HII region is extremely 
small, the
\lya photons produced behind the QSO (and emitted towards us) 
will encounter the edge of the
ionized bubble when it has grown up to a much larger scale, 
substantially reducing the damping
wing absorption.

\subsection{\lya emissivity from the I-front}

\begin{figure*}
\epsscale{0.9}
\plottwo{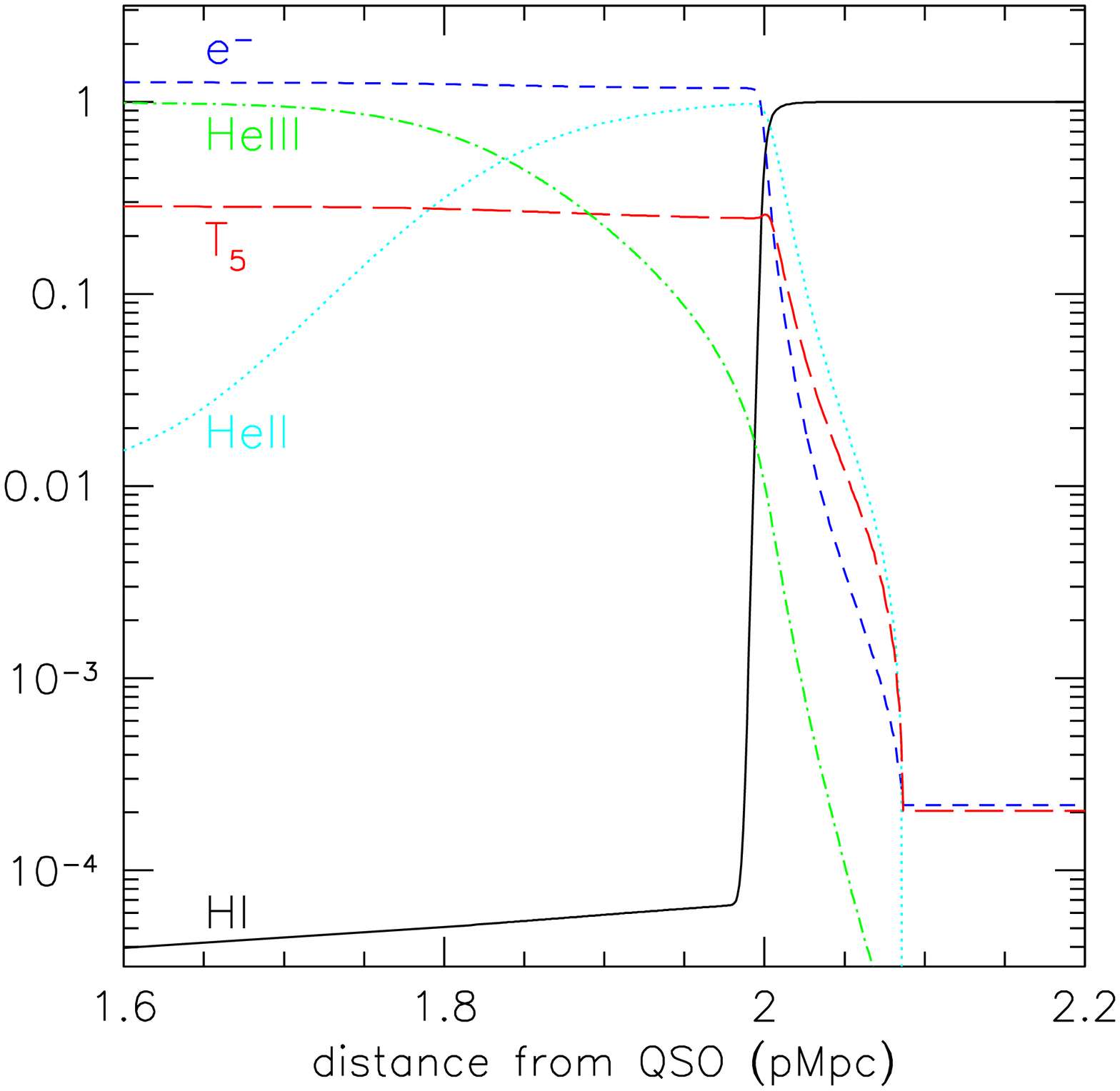}{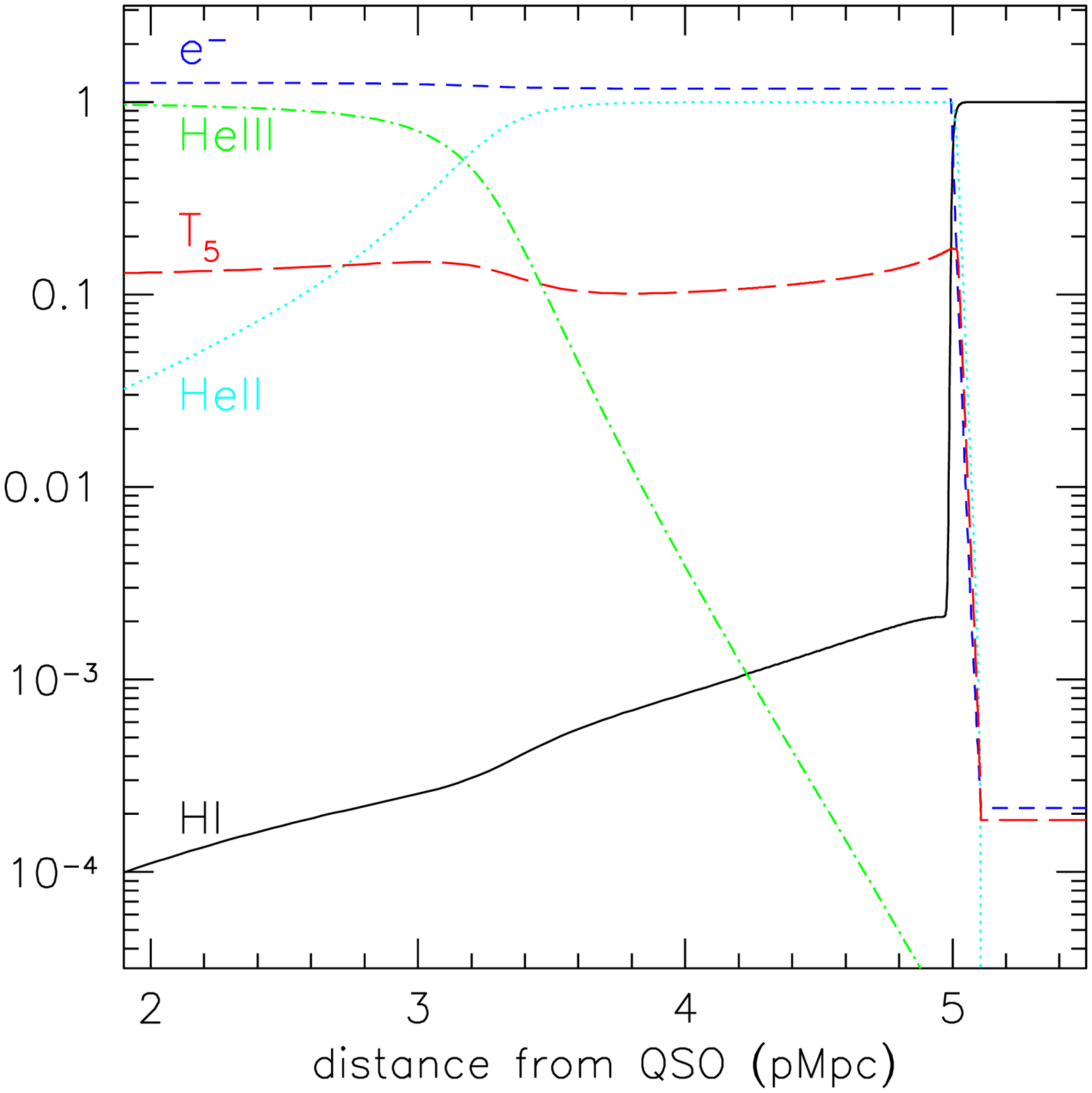}
\caption{Radial (rest-frame) temperature and ionization
profiles for our reference model at two different 
expansion epochs.
Solid, dotted and dash-dotted lines respectively show
the HI, HeII and HeIII fractions.
The long-dashed line indicates 
the temperature expressed in units of $T_5=(T/10^5\,\mathrm{K})$. 
The short-dashed line represents the
ratio $n_{\mr{e}}/n_{\mr{HI}}$ which is larger than 1 inside the HII region 
because of the contribution of partially or totally ionized Helium. 
Note that the scales and the x-ranges of the
two plots are very different.}
\label{f5} 
\end{figure*}

\begin{figure}
\plotone{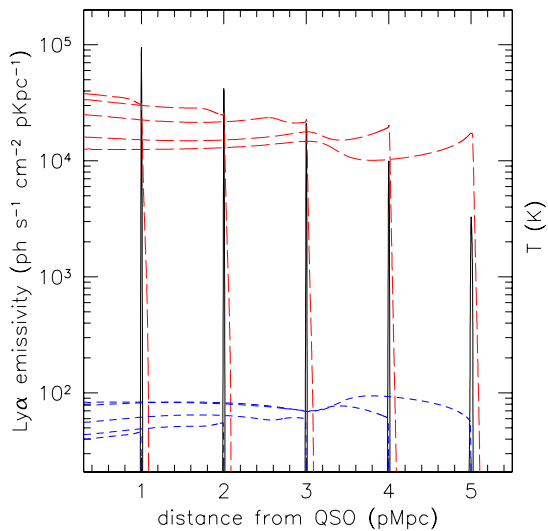}
\caption{\lya emissivity from collisional excitations (black solid line) 
and recombinations (blue short-dashed line) for our reference model at
different expansion epochs of the QSO I-front. 
For reference, we also show the evolution of the
temperature profiles (red long-dashed line).}
\label{f6} 
\end{figure}

Using the time evolution of the I-front derived in the previous section,
we want now to compute the \lya signal produced  
at given distance (or, equivalently, QSO age) from the central source. 

In Figure \ref{f5} 
we show the temperature and ionization profiles for our reference model
at two different epochs in the QSO rest-frame.
At early times (left panel), the HII/HI transition
is very sharp and the temperature (red, long-dashed line) 
slowly decreases going outwards until it suddenly 
drops down in the neutral region. 
In particular,
there is a small peak in the temperature profile at the position of the 
I-front ($x_{\mr{HI}}\sim0.5$, where $T\sim3\times10^4$ K) 
due to the radiative transfer effects that increase
the hardness of the ionizing spectrum. 
The HeII region extends outward of the HII zone, 
contributing to the free electron density (blue dotted line)
and to the temperature. 
At later stages (right-panel), the HII/HI transition 
is less sharp and 
the inner HII region is colder, $T\sim10^4$ K. 
However, the temperature peak at the I-front is 
more pronounced, reaching a value of $T\sim2\times10^4$ K. 
The HeIII region is now well inside the 
HII zone, with an important effect on the temperature profile (note the 
temperature drop at $\sim3$ pMpc).

In Figure \ref{f6} we show 
the evolution of the \lya emissivity from collisional 
excitations (black solid line), 
and radiative recombinations (blue, short-dashed line). 
As expected, 
\lya photons 
generated by collisional excitations are only produced on the I-front,
while those produced by radiative recombinations are emitted within
the entire HII region.
Photons from collisional excitations dominate the total \lya emissivity,
but their contribution diminishes with time
as a consequence of the decreasing I-front temperature
(red long-dashed line).

\subsection{Ly$\alpha$ Radiative Transfer}

\begin{figure*}
\epsscale{1.05}
\plottwo{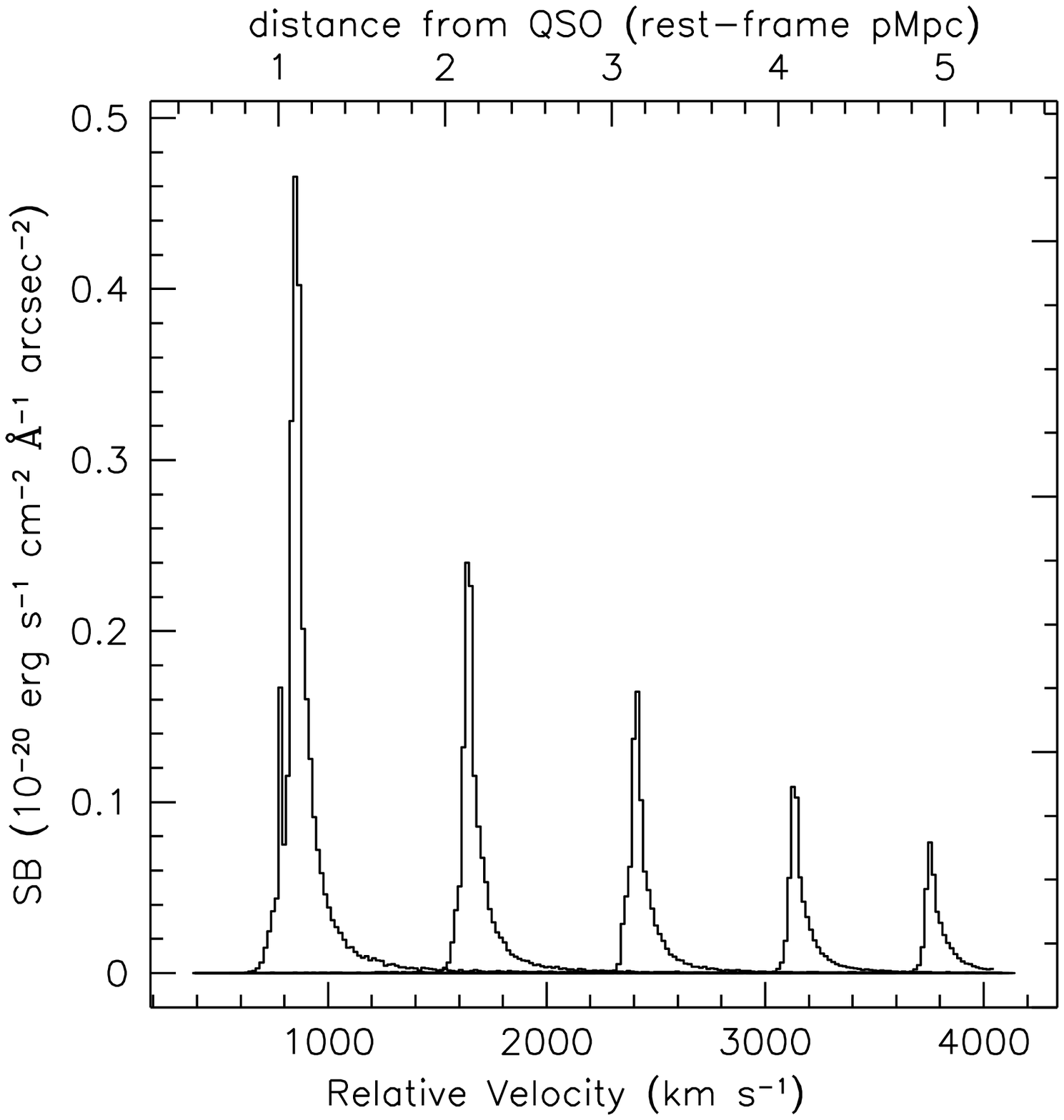}{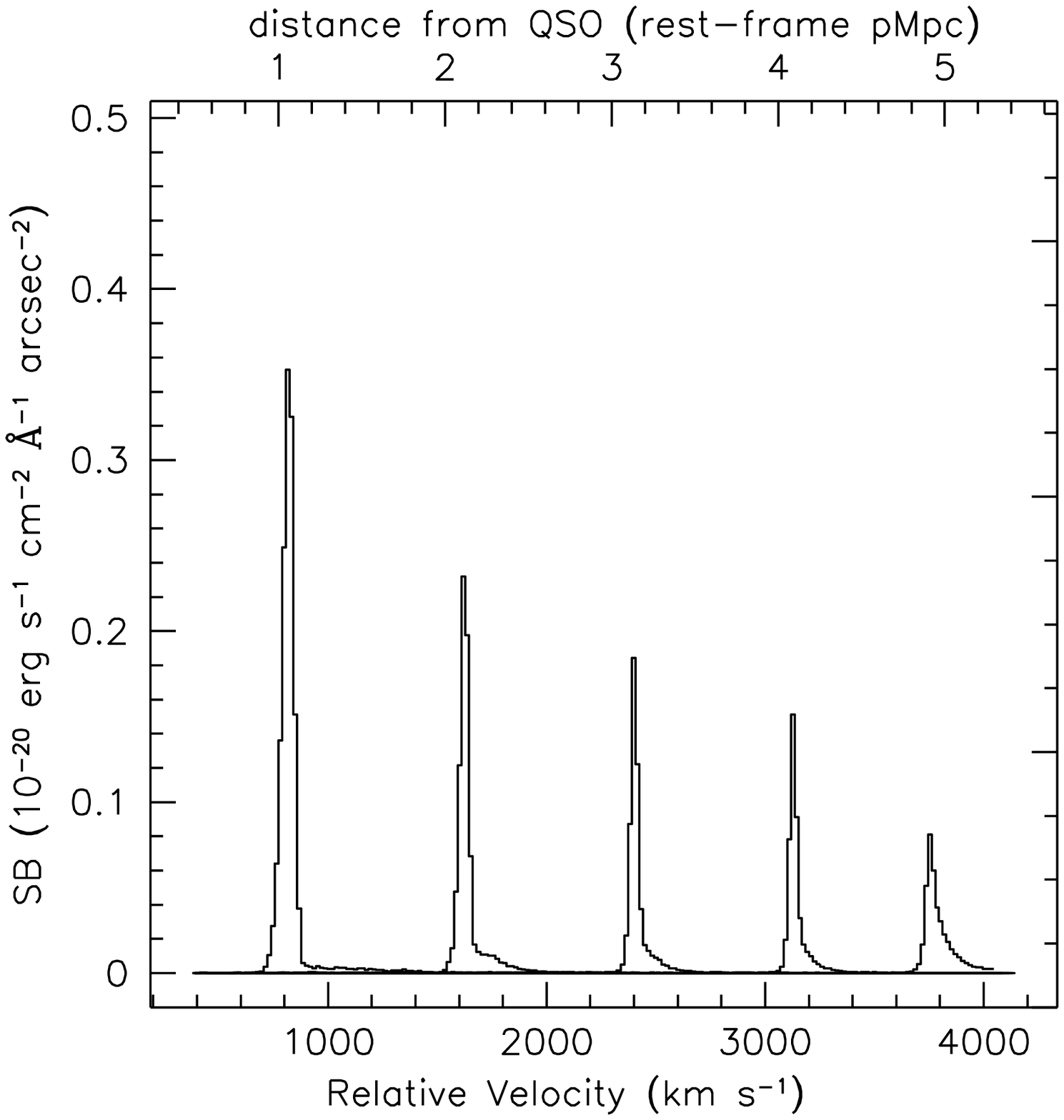}
\caption{Spectral evolution of the \lya emission from the expanding 
I-front (behind the QSO) for our reference model.
The different lines corresponds to 
the five different epochs for which the emissivities are shown in 
Figure \ref{f6}.
The spectra have been obtained with our \lya RT code
for a non-relativistic front (left) and including relativistic expansion (right).
Absorption by (residual) HI along the line of sight is included.}
\label{f7}
\end{figure*}

The density of neutral hydrogen within the I-front is high enough to make the
medium extremely optically-thick 
($\tau_{\lyam}\gtrsim10^4$ at the line center) to the \lya photons generated 
via collisional excitations.
Moreover, the residual neutral hydrogen along the line of sight 
may substantially change the observed \lya line shape and flux. 
A numerical treatment of the transfer of resonant line radiation is
required to fully account for these effects.
We thus compute
the observed properties of the \lya emission from both 
collisional excitations and radiative recombinations 
using an updated version of   
the three-dimensional \lya Monte Carlo code 
presented in Cantalupo et al. (2005). 
In particular, the new version includes two improvements.
First, it uses a more precise (and computationally less expensive) 
analytical fit of the Voigt-Hijertig function proposed by Tepper-Garcia (2006).
Second, it performs the line transfer in a medium
with a spatially varying temperature.
The presence of temperature gradients
not only changes
the \lya emissivity but also the shape of the absorption cross-section and 
thus the scattering process.
In particular, a \lya photon generated in a region with high temperature can 
more easily penetrate a colder zone before getting eventually absorbed. 
Thus, the photons produced in the (warm) I-front will diffuse more
in the neutral (cold) region than expected for a single-temperature medium. 

Given the small thickness of the I-front compared with the radius of the HII 
region, we can work in the plane-parallel approximation.
In the absence of Hubble expansion and assuming no absorption from
residual HI in the ionized region, the \lya spectrum emerging from the I-front
would be characterized by the symmetric, double-peaked emission 
typical of plane-parallel, uniform-density media (e.g., Neufeld 1990).
In this case, \lya photons would be emitted with a cosine law
over a net solid angle of $\pi$ (see e.g. Gould \& Weinberg 1996).
However, 
both the Hubble expansion and the presence of residual HI
increase the number of scatterings for the photons in 
the blue peak which are then 
absorbed and re-emitted over a wider solid angle 
and frequency interval.
The net effect is a suppression of the blue peak in the spectrum.
\begin{footnote}{This neglects peculiar velocities that, in general, 
modify the spectral shape of either the red or the blue peak, as 
shown in Cantalupo et al. 2005.}
\end{footnote}
The strength of this suppression depends on the residual HI fraction 
and, therefore, on the I-front distance from the QSO
(or, equivalently, on the QSO age). 
In most cases, the HI residual fraction is
high enough to erase the blue peak. 
On the other hand,
the damping wing absorption from HI lying outside the ionized region 
only reduces the \lya line flux by a few per cent at all epochs (without 
altering the line-shape). 

In the left panel of Figure \ref{f7}, 
we show the evolution of the observed \lya spectra 
obtained from the 
I-front profiles and emissivities presented in Figure \ref{f6}. 
At early stages, the blue peak 
is still detectable but, in general,
it is strongly suppressed. 
Therefore, the emission consists of a single redshifted 
peak with an asymmetric tail on the red side
(i.e. towards the direction of the cold neutral gas). 
The broadening of the observed peak is
significant compared with the sharpness of the emissivity in Figure \ref{f6}.
Both the suppression of 
the blue-peak and the damping-wing absorption from 
external HI reduce the integrated surface brightness with respect to
the static, plane-parallel case. 
Our simulation suggests that the observable integrated SB roughly 
corresponds to
58 per cent of the ideal case (solid line in Figure \ref{f8}).

\subsection{\lya Radiative Transfer and Relativistic I-fronts}

In the previous section we treated the I-front as static, i.e. we
assumed that the
\lya photons leave the I-front on a time scale which is short
with respect to the characteristic ionization time of the IGM.  
This assumption, however, is not valid 
for young QSOs
when the I-front expansion is ultra-relativistic.
In this case, as the \lya photons scatter, the medium becomes optically thinner and the 
optically-thick edge of the I-front moves outward. 
If the I-front speed is close to the speed of light,
the \lya photons find themselves inside the HII region after few scatterings. 
Once there,
they will be scattered over a wider solid angle ($4\pi$ at maximum) 
with respect to the semi-infinite slab case.
Therefore, ultra-relativistic I-fronts should emit 
a \lya flux which is reduced by up to
a factor of $4$ with respect to the static plane-parallel emission
and up to a factor of $2$ 
with respect to the non-relativistic case discussed in the previous section.

\begin{figure}
\plotone{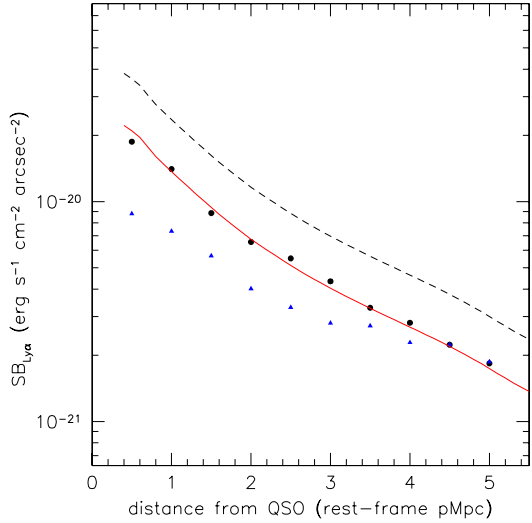}
\caption{Integrated \lya SB of the expanding I-front (behind the QSO) of our reference model at different expansion epochs,
as obtained by our \lya RT simulations (solid circles). The dashed line represent the expected SB from 
a plane-parallel (and optically thick) medium given by the integrated \lya emissivity inside the I-front
 ($SB^{pp}_{\lyam}$). The solid line represent $0.58\times SB^{pp}_{\lyam}$. Finally, the solid triangles shows the result
of our \lya RT that includes the effect of the I-front relativistic expansion (see text for details).}
\label{f8}
\end{figure}

A full treatment of this effect would require to perform the
\lya RT on a time-evolving grid, 
with a significant increase of the algorithm complexity and of the requested 
computational time. 
However, the fact that the I-front profile shifts in a self-similar way
on the relevant time scales gives us the possibility to properly treat \lya 
scattering in these extreme conditions.
We thus perform the \lya RT in the rest-frame of the I-front
(where the HI density is nearly time-independent). 
In this reference frame, \lya photons are preferentially emitted 
in the direction opposite to the propagation of the I-front.
The resulting time evolution of the line shape and of the integrated
SB are shown in the right panel
of Figure \ref{f7} and in Figure \ref{f8}, respectively.
Both the integrated SB and the line broadening 
are decreased by a factor $\sim2$ during the early stages of
the I-front expansion, while they 
coincide with the non-relativistic results at later epochs.
Note that the asymmetry of the line profile is reversed (skewed towards
the blue side) at very early times.

\section{Discussion}

\begin{figure}
\epsscale{1.1}
\plotone{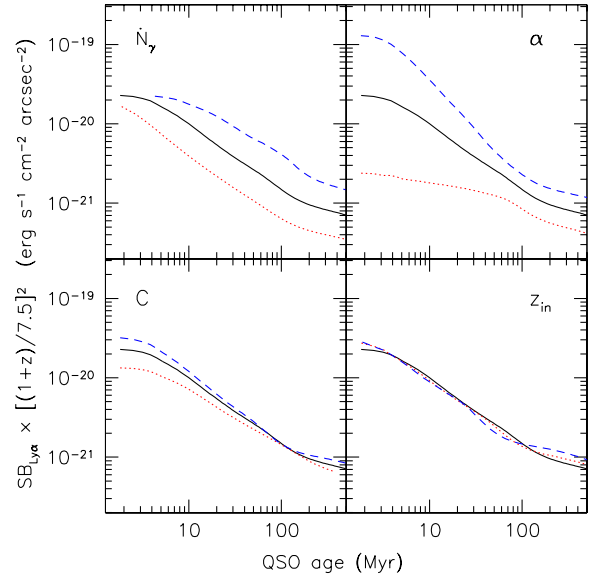}
\caption{
Expected \lya Surface Brightness 
produced by the QSO I-front as it crosses an initial neutral region of the IGM at mean density. 
In each panel, the reference model (i.e, total ionizing rate $\dot{N_{\gamma}}=10^{57}$ ph s$^{-1}$, 
far-UV spectral index $\alpha=-1.7$, initial redshift $z_{in}=6.5$, clumping factor $C=35$) 
is represented by the black solid line. The effect of the variation of one parameter with respect to the reference model 
is shown in each panel with the red dotted and the blue dashed lines representing, respectively: 
$\dot{N}_{\gamma}=10^{56}$ and $\dot{N}_{\gamma}=10^{58}$ (upper-left panel);
$\alpha=-2.5$ and $\alpha=-1.0$ (upper-right panel); 
$C=20$ and $C=50$ (bottom-left panel); $z_{in}=8$ and $z_{in}=10$ (bottom-right panel).
 The corresponding SB for a region with initial neutral fraction $x_{\mr{HI}}$ and overdensity
$(1+\delta)$ will scale approximatively like $x^2_{HI}(1+\delta)^{1/2}$ (see text for details).
 Note that the observed SB at very early stages (e.g., $t_{\mr{Q}}\lesssim5$ Myr for the reference model) may be substantially
lower than the values shown in the figure because of the high-relativistic expansion of the I-front (see text for details).
 }\label{fD}
\end{figure}

The parameters of the ionizing source in
our reference model have been chosen based on observations of
QSOs showing a GP trough at $z\sim 6$.
In this case, the \lya SB produced by the I-front which propagates 
within a neutral patch of the IGM at mean density
always lies between $10^{-21}-10^{-20}$ erg s$^{-1}$ cm$^{-2}$ arcsec$^{-2}$ 
(see Figure \ref{f8}).
How does the \lya SB change with varying the QSOs and the surrounding medium properties? 
Moreover, what is the redshift dependence of our results?
In order to answer these questions, we performed a series of simulations in which we kept fixed all
the parameters of the reference model but one.
 The resulting SBs (for a neutral patch of the IGM at mean density) are shown in Figure \ref{fD}.
In each panel, we indicate on the top-left the varying parameter (see caption for details) and 
the reference model is always represented by the black solid line.
The SB is obtained from the integrated emissivity and corrected for absorption and
RT effects, as described in \S 4. The correction for the 
relativistic I-front expansion in the \lya RT is not included. However, as shown in \S 4.6, the
decrease in the integrated SB is at maximum a factor of 2 when the expansion is ultra-relativistic and
it is negligible at later stages. 

 From the top-left panel in Figure \ref{fD}, we notice that, for a given QSO age, $t_{\mr{Q}}$, the expected
\lya SB is roughly proportional to $\dot{N}_{\gamma}^{1/3}$, unless the I-front expansion 
is still ultra-relativistic (i.e. $t_{\mr{Q}}\lesssim 5$ Myr for our reference model). 
Note that this corresponds to a linear scaling with the ionizing flux at the I-front position.
After the ultra-relativistic period,
and before recombinations start to play a significant role, 
the time evolution of the \lya SB is well represented by a power law (SB $\propto t_{\mr{Q}}^{-1}$)
independently of the QSO luminosity.
 The slope of the power law is mainly determined by the spectral index of the QSO spectrum 
(see top-right panel in Figure \ref{fD}), as expected by the strong temperature dependence
of the collisional-excitation rate. In particular, we found that the SB evolves proportionally
to $t_{\mr{Q}}^{-(\alpha+2.7)}$. 
Varying the clumping factor ($20<C<50$) 
has little effect on the SB evolution (bottom-left panel in Figure \ref{fD}). 
This is because the increase in the collisional excitation rate due to a more clumpy medium
is partially balanced by the fact that the medium becomes colder (mainly because of the
cooling from collisional excitations themselves). 
This self-regulation is more efficient as the I-front slows down. 

Finally, in the bottom-right panel of Figure \ref{fD}, we show that
the expected SB$_{\lyam}$, for a given $t_{\mr{Q}}$, decreases proportionally to $(1+z)^{-2}$ (instead of
$(1+z)^{-1}$ as expected for a medium with fixed temperature, see \S 3.1). 
Note that, in order to evidence this relation, the y-axis of the figure has been properly 
normalized. 

In the models discussed so far we only considered the \lya SB generated within a neutral patch
of the IGM at mean density. As we already shown in \S 3, 
the expected emission is proportional to the square of the initial neutral fraction ($x_{\mr{HI}}$) 
(i.e., the medium neutral fraction before the arrival of the I-front).
In order to estimate how the expected SB changes with the local overdensity 
($\delta\equiv(\rho-\rho_0)/\rho_0$), 
we performed a series of simulations placing slabs with different values of $(1+\delta)$ at a given distance
from the QSO. We found that, in overdense regions ($\delta\lesssim50$),
the expected \lya SB is nearly proportional to $(1+\delta)^{1/2}$. 
In summary, assuming $20\lesssim C\lesssim 50$, 
the \lya SB produced by a QSO I-front is of the order of:
\begin{equation}
\begin{split}
SB_{\lyam }&\sim \ 10^{-20} \cdot x^{2}_{\mathrm{HI}} (1+\delta)^{1/2} 
                 \cdot \left[\frac{t_{\mr{Q}}}{10 \mathrm{Myr}}\right]^{-1} \\ 
	   &\times \left[\frac{\dot{N}_{\gamma}}{10^{57} \mathrm{s}^{-1}}\right]^{1/3} 
                  \left[\frac{1+z}{7.5}\right]^{-2}  \mathrm{erg}\ \mathrm{s}^{-1} 
                    \mathrm{cm}^{-2} \mathrm{arcsec}^{-2}\ , 
\end{split}
\label{SBdep}
\end{equation}
for $\alpha\simeq-1.7$, $5\lesssim t_{\mr{Q}}\lesssim 100$ Myr, and
$\delta\lesssim50$.
Note, however, that QSOs with harder ionizing spectra ($\alpha\simeq
-1$) produce significantly brighter I-fronts (blue dashed line in Figure 9).

We do not consider here the \lya SB produced within more overdense regions. In this case, 
a proper treatment should also consider hydrodynamical effects.
Fluorescent \lya emission from
overdense ($\delta>100$) 
regions could be as bright as the I-front emission at mean density. 
However, these dense regions will appear significantly more compact. 
An extensive study of the \lya emission produced by overdense regions 
and the effect 
of IGM inhomogeneities will be presented in a future paper (Cantalupo et al., in preparation).

\subsection{Uncertainties and limitations}

For very young QSOs, when the expansion-speed of the I-front ($v_{\mr I}$) is very close to the speed of light,
our model might overestimate the actual \lya emission.
In this case,
the photo-ionization time scale ($t_{\mr{ion}}$) at  the I-front 
(roughly given by $t_{\mr{ion}}\sim S/v_{\mr I}$, where $S$ is the I-front thickness) can be
shorter than the collisional-excitation time scale $t_{\mr{CE}}\sim(C n_{\mr{HI}}q(T))^{-1}$.
In our reference model, this effect is important for $t_{\mr{Q}}\lesssim 5$ Myr (and earlier at higher redshifts). 
In overdense regions the I-front is slower and collisional excitations are 
faster with respect to the IGM at mean density. Therefore, the \lya emission produced in these regions
is less sensitive to this effect. 
On the other hand, I-fronts expanding around young quasars 
lie very close to the source. For instance, the 
I-front behind a bright QSO at $z\sim6.5$ with $t_{\mr{Q}}\sim5$ Myr appears at 
a proper distance of 1 Mpc (see Figure 4). If the opening angle of the QSO emission cone 
is $45^o$ (with respect of the line of sight), the 
total projected area covered by the I-front will extend over $\sim12$ arcmin$^2$ 
(or $\sim 60$ arcmin$^2$ if the quasar emits isotropically).
Therefore,  older QSOs provide a larger projected area for detecting the 
I-front emission and a better constraint
on the properties of both the quasar emission and the surrounding IGM.
Moreover, older and luminous QSOs produce HII bubbles that are
larger than possible pre-existing ionized regions generated 
from clustered star-forming galaxies.

Our RT simulations do not consider X-ray radiation produced by the QSO.
However, X-ray photons travel unimpeded well 
beyond the position of the emitting I-front and do not modify the properties of the \lya signal. 
In the presence of pre-existing X-ray background, the initial gas temperature will be higher than  
assumed in our models ($T\sim100$ K) and this will slightly increase 
the \lya emission from collisional excitations (provided
that the medium is still significantly neutral before the arrival of the I-front).

For simplicity, we have assumed a steady ionizing emission rate for the quasars. We also did not account
for the possible contribution of other local sources that
can change the apparent shape of the QSO I-front and the properties of the 
surrounding medium (see e.g., Yu 2005; Yu \& Lu 2005; Wyithe \& Loeb 2007; Lidz et al. 2007). 
Addressing these effects 
requires performing high-resolution RT within a fully cosmological simulation and is beyond the scope of the present study.

\subsection{Detectability}

To date, eight QSOs have been detected at redshift $z>6.1$ (Fan et al. 2003, 2004; Willot et al. 2007) and all of
them have significant ($\Delta z>0.1$) dark absorption 
troughs in their spectra in proximity of the QSO redshift.
 Although this is not necessarily evidence for a surrounding  
neutral medium, these objects represent the best targets for detecting the I-front \lya emission.

Let us take, as a practical example, the case of QSO J1148+5251. This quasar has the 
most accurate redshift measurement($z=6.419\pm0.001$, Walter et al. 2003) and shows complete 
Ly$\alpha$-Ly$\beta$ troughs corresponding to
an apparent proximity region of $R_{\mr s}\sim5$ physical Mpc (Wyithe et al. 2004, Yu \& Lu 2005).
 Assuming that the medium surrounding the QSO was still significantly neutral ($\langle x_{\mr{HI}}\rangle > 0.1$) 
before the QSO turned on, from the estimated ionizing rate  
($\dot{N}_{\gamma}\sim 2\times 10^{57}$ ph s$^{-1}$, see e.g. Yu \& Lu 2005) we 
expect $8\lesssim t_{\mr{Q}}\lesssim 30$ Myr (see Figure \ref{f4}).
 This corresponds, for a QSO spectral index $\alpha=-1.7$, to
$0.5\lesssim SB_{\lyam}\lesssim 1.5\times 10^{-20}$ \ergscmarcsec (see equation (\ref{SBdep}) 
and Figure \ref{fD}) for   
a neutral patch of the IGM at mean density. 

 Is this signal detectable? Although faint (about 3 orders of magnitude below the 
sky background), the expected emission may extend over a projected area on the sky up to a few hundred square arcmin.
By means of moderately-high resolution spectroscopy (R=1000-3000) and integrating the signal over a 
fraction of the slit length, single neutral patches of the IGM (even at mean density) can be already
detected from the ground with current facilities and long integration times ($T\sim 40$ hr)
if they extends over few arcminutes scales (or smaller for slightly overdense regions).
For instance, the expected signal-to-noise (S/N) ratio 
corresponding to the above example and for a ground-based
observation (in the atmospheric window at $\sim0.9\mu m$) 
will be of the order of:
\begin{equation}
\begin{split}
S/N   &\sim 7\times  C_{\rm f} \left[\frac{D}{8\ \mathrm{m}}\right] \left[\frac{\zeta}{0.8}\right]  
       \left[\frac{f}{0.25}\right]^{1/2} \\
    & \left[\frac{\Delta_s}{3\mathrm{\AA}}\right]^{-1/2}
       \left[\frac{T}{40\ \mathrm{hr}}\right]^{1/2} \left[\frac{\Delta\Omega}{180\ \mathrm{arcsec}^2}\right]^{1/2}\ , 
\end{split}
\end{equation}
 where $C_{\rm f}$ is the slit covering factor (i.e. the fraction of the slit where the \lya emission is present),
 $D$ is the telescope diameter, $\zeta$ the atmospheric transmission, $f$ the system efficiency,
 $\Delta_s$ the spectral bin, $T$ the integration time and $\Delta\Omega$ the area of the sky covered
by the slit. For a slit width of $1''$, the above $S/N$ implies that we can detect
a single neutral patch of the IGM (i.e., $C_{\rm f}=1$) at mean density if it has at least 
a linear size of $l\sim3'$ (corresponding to $\sim1$ pMpc at $z\sim6.4$).

 Even if single neutral patches cannot be detected,
a proper integration over the total slit length (that can be substantially increased, for instance, using
a multi-slit plus filter technique, see e.g., Cantalupo et al. 2007) 
may reveal at least the position of the \lya emitting I-front.
 Once the I-front has been located, we can use its apparent distance from the QSO 
to determine the QSO age and to constrain the value of $\langle x_{\mr{HI}}\rangle$
(see Figure 4).
Possible limitations are given by the unknown opening angle of the QSO emission (that can reduce
the projected area over which the I-front may extend) and the difficulties of
subtracting the sky lines.  

A significant improvement is expected from the next generation of space telescopes like JWST,
although the background (i.e., the Zodiacal light) will still be 2 orders of magnitude 
brighter than the expected signal.
With JWST, we will be able to detect single neutral patches of the IGM on smaller scales 
than allowed from the ground.
Moreover observations will not be limited to the narrow redshift ranges permitted
by the atmospheric windows. 
For a given QSO age, the \lya SB decreases proportionally to $(1+z)^2$ (instead of the usual $(1+z)^4$).
The next generation of space telescopes will thus be able
to map HI distribution during the EoR in a broad redshift range.

\section{Summary and Conclusions}

We have presented a new method to directly map the neutral hydrogen 
distribution during the reionization epoch, 
and to measure the age of the highest-redshift QSOs.
We have shown that collisional excitations are an
efficient mechanism to produce \lya photons, even in the low density IGM, 
provided that nearly 50 per cent of the hydrogen is neutral and the
local temperature is as high as $T\gtrsim2\times10^4$ K.
These conditions are achieved within the I-fronts produced by luminous UV
sources, like QSOs, as they expand into the surrounding IGM.
The observable \lya photons are
those emitted by the I-front \emph{behind} the QSO 
(with respect to the observer).
These photons can cross the large HII region lying in front of
their emission point without being significantly scattered.
The emerging signal
traces the HI distribution at the location of the QSO I-front
(similarly to future 21-cm tomography) since
the expected SB
roughly scales as the square of the (initial) local  
HI fraction.
The angular distribution of the \lya emission and its
distance from the QSO
constrains both the properties of the source 
(i.e. the QSO opening angle and age) 
and of the surrounding medium (i.e., the 
average neutral fraction).
 
Using detailed radiative transfer simulations that include finite light-speed
effects, we have shown that the expected  emission appears as a 
single (broad) line with a width of
$100-200$ km s$^{-1}$.
The \lya SB of a typical QSO I-front
that propagates at $z_{\rm Q}=6.5$ within
a fully neutral patch of the IGM at mean density 
lies in the range 
${\mr SB}_{\lyam}\sim10^{-21}-10^{-20}$ erg s$^{-1}$ cm$^{-2}$ arcsec$^{-2}$.
QSOs with very hard spectra ($\alpha\sim-1$) produce a significantly brighter 
signal at early phases ($t_{\mr{Q}}\lesssim10$ Myr). 
Interestingly, for a given QSO age,
the \lya SB of the I-front scales as $(1+z_{\rm Q})^{-2}$.

The signal from neutral patches of the IGM extending over a few arcmin scales
is already detectable by current ground facilities
with the use of moderately/high resolution spectroscopy.
The next generation of space and ground based telescopes will allow us
to draw high signal-to-noise maps of the HI distribution with higher 
angular resolution.
Combining this information with
a measure of the I-front position, 
we will be able to simultaneously constrain the reionization history 
and the emission properties of the highest-z QSOs.

\acknowledgements
SC is supported by the Swiss National Science Fundation.


\begin{thebibliography}{}

\bibitem[Abel \& Haehnelt(1999)]{1999ApJ...520L..13A} Abel, T., \& 
Haehnelt, M.~G.\ 1999, \apjl, 520, L13 

\bibitem[Baker \& Menzel(1938)]{1938ApJ....88...52B} Baker, J.~G., \& 
Menzel, D.~H.\ 1938, \apj, 88, 52 

\bibitem[Becker et al.(2001)]{2001AJ....122.2850B} Becker, R.~H., et al.\ 
2001, \aj, 122, 2850 

\bibitem[Bolton \& Haehnelt(2007)]{2007MNRAS.374..493B} Bolton, J.~S., \& 
Haehnelt, M.~G.\ 2007, \mnras, 374, 493 

\bibitem[Cantalupo et al.(2005)]{2005ApJ...628...61C} Cantalupo, S., 
Porciani, C., Lilly, S.~J., \& Miniati, F.\ 2005, \apj, 628, 61 

\bibitem[Cantalupo et al.(2007)]{2007ApJ...657..135C} Cantalupo, S., Lilly, 
S.~J., \& Porciani, C.\ 2007, \apj, 657, 135 

\bibitem[Fan et al.(2002)]{2002AJ....123.1247F} Fan, X., Narayanan, V.~K., 
Strauss, M.~A., White, R.~L., Becker, R.~H., Pentericci, L., \& Rix, H.-W.\ 
2002, \aj, 123, 1247 

\bibitem[Fan et al.(2003)]{2003AJ....125.1649F} Fan, X., et al.\ 2003, \aj, 
125, 1649 

\bibitem[Fan et al.(2006)]{2006AJ....132..117F} Fan, X., et al.\ 2006, \aj, 
132, 117 

\bibitem[Fan et al.(2006)]{2006ARA&A..44..415F} Fan, X., Carilli, C.~L., \& 
Keating, B.\ 2006, \araa, 44, 415 

\bibitem[Furlanetto et al.(2004)]{2004ApJ...613....1F} Furlanetto, S.~R., 
Zaldarriaga, M., \& Hernquist, L.\ 2004, \apj, 613, 1 

\bibitem[Furlanetto et al.(2006)]{2006PhR...433..181F} Furlanetto, S.~R., 
Oh, S.~P., \& Briggs, F.~H.\ 2006, \physrep, 433, 181 

\bibitem[Giovanardi et al.(1987)]{1987A&AS...70..269G} Giovanardi, C., 
Natta, A., \& Palla, F.\ 1987, \aaps, 70, 269 

\bibitem[Gnedin \& Ostriker(1997)]{1997ApJ...486..581G} Gnedin, N.~Y., \& 
Ostriker, J.~P.\ 1997, \apj, 486, 581 

\bibitem[Gould \& Weinberg(1996)]{1996ApJ...468..462G} Gould, A., \& 
Weinberg, D.~H.\ 1996, \apj, 468, 462 

\bibitem[Gunn \& Peterson(1965)]{1965ApJ...142.1633G} Gunn, J.~E., \& 
Peterson, B.~A.\ 1965, \apj, 142, 1633 

\bibitem[Haiman(2002)]{2002ApJ...576L...1H} Haiman, Z.\ 2002, \apjl, 576, 
L1 

\bibitem[Hogan \& Weymann(1987)]{1987MNRAS.225P...1H} Hogan, C.~J., \& 
Weymann, R.~J.\ 1987, \mnras, 225, 1P 

\bibitem[Iliev et al.(2006)]{2006MNRAS.371.1057I} Iliev, I.~T., et al.\ 
2006, \mnras, 371, 1057 


\bibitem[Lidz et al.(2007)]{2007astro.ph..3667L} Lidz, A., McQuinn, M., 
Zaldarriaga, M., Hernquist, L., \& Dutta, S.\ 2007, ArXiv Astrophysics 
e-prints, arXiv:astro-ph/0703667 

\bibitem[Kashikawa et al.(2006)]{2006ApJ...648....7K} Kashikawa, N., et 
al.\ 2006, \apj, 648, 7 

\bibitem[Madau \& Rees(2000)]{2000ApJ...542L..69M} Madau, P., \& Rees, 
M.~J.\ 2000, \apjl, 542, L69 

\bibitem[Madau et al.(1997)]{1997ApJ...475..429M} Madau, P., Meiksin, A., 
\& Rees, M.~J.\ 1997, \apj, 475, 429 

\bibitem[Malhotra \& Rhoads(2006)]{2006ApJ...647L..95M} Malhotra, S., \& 
Rhoads, J.~E.\ 2006, \apjl, 647, L95 

\bibitem[Martin(1988)]{1988ApJS...66..125M} Martin, P.~G.\ 1988, \apjs, 66, 
125 

\bibitem[Maselli et al.(2007)]{2007MNRAS.376L..34M} Maselli, A., Gallerani, 
S., Ferrara, A., \& Choudhury, T.~R.\ 2007, \mnras, 376, L34 

\bibitem[McQuinn et al.(2007)]{2007arXiv0704.2239M} McQuinn, M., Hernquist, 
L., Zaldarriaga, M., \& Dutta, S.\ 2007, ArXiv e-prints, 704, 
arXiv:0704.2239 

\bibitem[Mesinger \& Haiman(2004)]{2004ApJ...611L..69M} Mesinger, A., \& 
Haiman, Z.\ 2004, \apjl, 611, L69 

\bibitem[Miralda-Escude \& Rees(1994)]{1994MNRAS.266..343M} Miralda-Escude, 
J., \& Rees, M.~J.\ 1994, \mnras, 266, 343 

\bibitem[Miralda-Escude \& Rees(1998)]{1998ApJ...497...21M} Miralda-Escude, 
J., \& Rees, M.~J.\ 1998, \apj, 497, 21 

\bibitem[Neufeld(1990)]{1990ApJ...350..216N} Neufeld, D.~A.\ 1990, \apj,
350, 216

\bibitem[Oh \& Furlanetto(2005)]{2005ApJ...620L...9O} Oh, S.~P., \& 
Furlanetto, S.~R.\ 2005, \apjl, 620, L9 

\bibitem[Osterbrock(1989)]{1989agna.book.....O} Osterbrock, D.~E.\ 1989,  
University Science Books, 1989

\bibitem[Page et al.(2007)]{2007ApJS..170..335P} Page, L., et al.\ 2007, 
\apjs, 170, 335 

\bibitem[Pengelly(1964)]{1964MNRAS.127..145P} Pengelly, R.~M.\ 1964, 
\mnras, 127, 145 

\bibitem[Santos(2004)]{2004MNRAS.349.1137S} Santos, M.~R.\ 2004, \mnras, 
349, 1137 

\bibitem[Shapiro \& Giroux(1987)]{1987ApJ...321L.107S} Shapiro, P.~R., \& 
Giroux, M.~L.\ 1987, \apjl, 321, L107 

\bibitem[Shapiro et al.(2006)]{2006ApJ...648..922S} Shapiro, P.~R., Iliev, 
I.~T., Alvarez, M.~A., \& Scannapieco, E.\ 2006, \apj, 648, 922 

\bibitem[Songaila(2004)]{2004AJ....127.2598S} Songaila, A.\ 2004, \aj, 127, 
2598 

\bibitem[Spergel et al.(2007)]{2007ApJS..170..377S} Spergel, D.~N., et al.\ 
2007, \apjs, 170, 377 

\bibitem[Taniguchi et al.(2005)]{2005PASJ...57..165T} Taniguchi, Y., et 
al.\ 2005, \pasj, 57, 165 

\bibitem[Telfer et al.(2002)]{2002ApJ...565..773T} Telfer, R.~C., Zheng, 
W., Kriss, G.~A., \& Davidsen, A.~F.\ 2002, \apj, 565, 773 

\bibitem[Tepper-Garc{\'{\i}}a \& 
Fritze-v.~Alvensleben(2006)]{2006MNRAS.369.2025T} Tepper-Garc{\'{\i}}a, T., 
\& Fritze-v.~Alvensleben, U.\ 2006, \mnras, 369, 2025 

\bibitem[Tran et al.(2004)]{2004ApJ...612L..89T} Tran, K.-V.~H., Lilly, 
S.~J., Crampton, D., \& Brodwin, M.\ 2004, \apjl, 612, L89 

\bibitem[Walter et al.(2003)]{2003Natur.424..406W} Walter, F., et al.\ 
2003, \nat, 424, 406 

\bibitem[White et al.(2003)]{2003AJ....126....1W} White, R.~L., Becker, 
R.~H., Fan, X., \& Strauss, M.~A.\ 2003, \aj, 126, 1 

\bibitem[White et al.(2005)]{2005AJ....129.2102W} White, R.~L., Becker, 
R.~H., Fan, X., \& Strauss, M.~A.\ 2005, \aj, 129, 2102 

\bibitem[Willott et al.(2007)]{2007arXiv0706.0914W} Willott, C.~J., et al.\ 
2007, ArXiv e-prints, 706, arXiv:0706.0914 

\bibitem[Wyithe \& Loeb(2004)]{2004Natur.432..194W} Wyithe, J.~S.~B., \& 
Loeb, A.\ 2004, \nat, 432, 194 

\bibitem[Wyithe et al.(2005)]{2005ApJ...628..575W} Wyithe, J.~S.~B., Loeb, 
A., \& Carilli, C.\ 2005, \apj, 628, 575 

\bibitem[Wyithe \& Loeb(2007)]{2007MNRAS.374..960W} Wyithe, J.~S.~B., \& 
Loeb, A.\ 2007, \mnras, 374, 960 

\bibitem[Yu(2005)]{2005ApJ...623..683Y} Yu, Q.\ 2005, \apj, 623, 683 

\bibitem[Yu \& Lu(2005)]{2005ApJ...620...31Y} Yu, Q., \& Lu, Y.\ 2005, 
\apj, 620, 31 


\end{thebibliography}
\end{document}